\documentclass[twocolumn,a4paper,superscriptaddress,floatfix,longbibliography]{revtex4}
\usepackage[english]{babel}
\usepackage[utf8]{inputenc}
\usepackage{amsmath}
\usepackage{graphicx,epstopdf}
\usepackage[colorinlistoftodos]{todonotes}
\usepackage{amssymb,epsfig,color,textcase}
\usepackage{hyperref}
\usepackage{doi}

\begin{document}

\title{The Jahn-Teller effect and spin--orbit coupling: friends or foes?
}

\author{Sergey V.~Streltsov$^*$}
\affiliation{Institute of Metal Physics, S. Kovalevskoy St. 18, 620990 Ekaterinburg, Russia}
\affiliation{Department of theoretical physics and applied mathematics, Ural Federal University, Mira St. 19, 620002 Ekaterinburg, Russia}
\email{streltsov@imp.uran.ru}

\author{Daniel I. Khomskii}
\affiliation{II. Physikalisches Institut, Universit$\ddot a$t zu K$\ddot o$ln,
Z$\ddot u$lpicher Stra$\ss$e 77, D-50937 K$\ddot o$ln, Germany}

\date{\today}

\begin{abstract}
The Jahn-Teller effect is one of the most fundamental phenomena important not only for physics, but also for chemistry and material science. Solving the Jahn-Teller problem and taking into account strong electron correlations we show that quantum entanglement of the spin and orbital degrees of freedom via spin--orbit coupling strongly affects this effect. Depending on the number of $d$ electrons it may quench (electronic configurations $t_{2g}^2$, $t_{2g}^4$, and $t_{2g}^5$), partially suppress ($t_{2g}^1$) or in contrast induce ($t_{2g}^3$) Jahn-Teller distortions. Moreover, in certain situations, interplay between the Jahn-Teller effect and spin--orbit coupling promotes formation of the ``Mexican hat'' energy surface facilitating various quantum phenomena.
\end{abstract}

\maketitle

%%%%%%%%%%%%%%%%%%%%%%%%%%%%%%%%%%%%%%%%%%%%%%%%%%%%%
\section{Introduction}
In considering transition metal compounds, one very often meets the situation with orbital degeneracy. Classical examples are the systems with Cu$^{2+}$ or Mn$^{3+}$ ions in octahedral coordination as in high-temperature superconductors or colossal magnetoresistive manganates\cite{Imada1998}. In this case a plethora of very interesting and nontrivial effects follows: the famous Jahn-Teller (JT) effect, orbital ordering, strong coupling between orbital and spin degrees of freedom (e.g.\ the well-known Googenough--Kanamori--Anderson rules)\cite{Goodenough}, very nontrivial quantum effects (vibronic physics), up to the appearance of rotational quantization, conical intersections and even the famous geometric (Berry) phase -- which actually first appeared in the literature in 1975 in the context of the JT effect\cite{Longuet-Higgins1, Longuet-Higgins1975}, long before the famous paper by Sir M.~Berry\cite{Berry1984}.

Yet another group of phenomena, related to spin--orbit coupling (SOC), came to the forefront recently (although this interaction itself is known for more than 100 years). It is one of the most actively studied topics in condensed matter physics at present. It is the cornerstone of such phenomena as the anomalous Hall effect, spin Hall effect\cite{Hirsch1999}, Rashba coupling\cite{Bychkov1984}, skyrmion physics\cite{Fert2017}. It also leads to Kitaev physics in transition metal compounds\cite{Takagi2019}.  Last but not least, topological insulators in most cases are due to SOC\cite{Qi2011}. 

A very natural and, strangely enough, still very scarcely touched topic is the mutual interplay of these two phenomena: orbital degeneracy, specifically the JT effect, and SOC\null. After the first, very old studies, see e.g.\ \cite{Jahn1938,Opik1957,Moffitt1957,Bates1978}, this connection remained largely unexplored. One of the few known statements in this field is that the appearance of $j_{\it eff}=1/2$ state in Ir$^{4+ }$ ions ``kills'' the JT distortions: strong SOC  lifts the three-fold $t_{2g}$ degeneracy  of the $t_{2g}^5$ levels and leads to the formation of the $j_{\it eff}=1/2$ Kramers doublet, without any extra degeneracy. Only rarely is the JT physics mentioned in the present literature devoted to such systems; notable exceptions are the discussions of the JT effect in excited states of Ir compounds\cite{Plotnikova2016}, the discovery of the role of pseudo-JT effect on lattice distortions in Sr$_2$IrO$_4$ and orbital order in Ca$_2$RuO$_4$\cite{Liu2019a}, 
possible appearance of SU(4) ``Kugel--Khomskii''-type interaction in $d^1$ systems with strong SOC on honeycomb lattice\cite{Yamada2018}, and the experimental study of unusual type of distortion in $5d^1$ system K$_2$TaCl$_6$\cite{Ishikawa2019}.  But the general answer to the question formulated in our title: are the JT effect and SOC friends or foes? -- seems to remain unanswered. This question is very important, both from the general theoretical point of view and for the discussion of properties of many real materials, especially those with $4d$ and $5d$ elements. The present paper is aimed at filling this gap. Surprisingly enough,  we found out that the answer to this very simply-formulated question  is not at all straightforward, as it depends on the specific situation: for some cases, such as $d^4$ and $d^5$, but also $d^2$ configurations, strong SOC suppresses JT distortion, whereas in other cases, notably $d^3$, SOC promotes or generates the JT effect in the configuration which is usually considered as ``orbitally dead''.

%%%%%%%%%%%%%%%%%%%%%%%%%%
\section{Qualitative arguments.}
We consider in this paper the case of partially-filled $t_{2g}$ levels (typical for the $4d$ and $5d$ ions in the usual low-spin state), for which the orbital moment and SOC are not quenched. It is assumed that the crystal field splitting between $t_{2g}$ and $e_g$ levels, $\Delta=10Dq$, is large enough so that one can neglect the $t_{2g}$--$e_g$ mixing. The $t_{2g}$ electrons in principle interact and can be split by the JT coupling with doubly-degenerate lattice distortions $E_g =\{Q_3, Q_2\}$ (i.e.\ this is the $t \otimes E$ problem)\cite{Bersuker1989,Bersuker2006}, where $Q_3$ is a tetragonal distortion (for concreteness we take $Q_3>0$ for tetragonal elongation), and $Q_2$  is the orthorhombic mode.

\begin{figure}[t]
   \centering
  \includegraphics[width=0.4\textwidth]{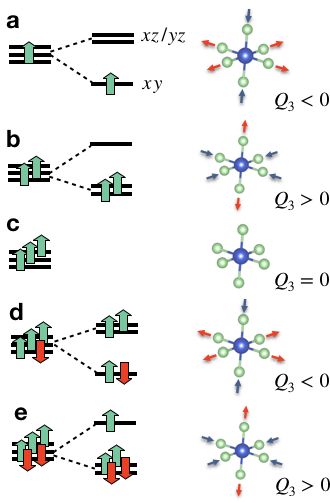}
  \caption{\label{Fig:CFS}  Diagrams of crystal-field splitting due to the Jahn-Teller effect without spin--orbit coupling for different number of electrons: (a)~$d^1$, (b)~$d^2$, (c)~$d^3$, (d)~$d^4$, and (e)~$d^5$.}
\end{figure}

For tetragonal distortions without SOC one has the situations illustrated in Fig.~\ref{Fig:CFS}(a)--(e). Here we distribute electrons according to the (first) Hund's rule, i.e.\ forming states with maximal possible spin, and show the corresponding level scheme  for the tetragonal JT distortion appropriate for each electronic configuration. One can see that indeed the situations shown in Fig.~\ref{Fig:CFS} are optimal from the point of view of both the Hund's coupling and the JT effect (crystal-field splitting). 

The JT effect tends to stabilize electrons on real (cubic) harmonics: $xy$, $xz$, $yz$, while SOC would favour occupation of {\it very  different} states, described by complex (spherical) harmonics, which are actually eigenstates of the effective orbital moment of the $t_{2g}$ triplet, $l_{\it eff} = 1$ (in what follows we will omit the subscript {\it eff}, but one has to remember that it is an effective moment)\cite{Sugano-book}: 
\begin{eqnarray}
| l^z_0 \rangle &=& | xy \rangle, \\
|l^z_{\pm 1} \rangle &=&  -\frac {1} {\sqrt 2} \left( {\rm i}|  xz \rangle \pm    |yz \rangle \right).
\label{lz1}
\end{eqnarray}
This is the essence of the physical effects considered in our paper: Transformation of the electronic wavefunctions due to SOC results in modification of the JT distortions, since the system aims to gain maximal total energy from both terms.

\section{Model}
We want to stress that the strategy chosen in the present paper is very different from theoretical approaches used previously. We do not use any perturbation theory and do not consider the formalism of double groups (see e.g.\ Ref.~\cite{Moffitt1957}), but solve the many-electron problem numerically exactly with the chosen form of the electron--lattice coupling. This allows us to take into account all many-particle effects such as strong on-site Hubbard repulsion, intra-atomic Hund's exchange and spin--orbit coupling. Thus, we treat numerically exactly the simplest (but physically very rich) situation of an isolated JT centre with partially filled $t_{2g}$ orbitals and consider the following Hamiltonian:
\begin{eqnarray}
\label{H}
\hat H = \hat H_{SOC} + \hat H_{JT} + \hat H_{U}.
\end{eqnarray}
The SOC must be treated in a full vector form as 
\begin{eqnarray}
\hat H_{SOC} = -\zeta \sum_{\alpha} \hat {\bf l}_{\alpha} \cdot \hat {\bf s}_{\alpha}
\end{eqnarray}
to preserve full rotational invariance. Here, $\hat {\bf l}_{\alpha}$ and $\hat {\bf s}_{\alpha} $ are orbital and spin moments of the $\alpha$-th electron, $\zeta$ is the SOC constant and the minus sign is due to the fact that we are working with the $t_{2g}$ orbitals having effective orbital moment $l_{\it eff}=1$\cite{Abragam}.  In the LS coupling scheme (for SOC weaker than the Hund's interaction) one can also write this part of the Hamiltonian as $H_{SOC} = -\lambda \hat {\bf L} \cdot \hat {\bf S}$, where $\hat {\bf L} =\sum_{\alpha} \hat  {\bf l}_{\alpha}$, $\hat {\bf S} =\sum_{\alpha} \hat {\bf s}_{\alpha}$ are the total orbital and spin moments of a particular configuration, and $\lambda = \zeta/2S$. 

The JT term includes the elastic energy and linear coupling of the electronic subsystem, described by projections of the angular moments $\bf l_{\alpha}$, with the two lattice distortions, characterized by normal coordinates $Q_2$ and $Q_3$ with respective coefficients $g$ and $B$~\cite{KK-UFN}: 
 \begin{eqnarray}
\hat H_{JT} &=& -\frac g{\sqrt 3} \sum_{\alpha} \left[  \left(\hat l^x_{\alpha}\right)^2 
 - \left( \hat l^y_{\alpha}\right)^2  \right] Q_2 \nonumber \\
 &-& g \sum_{\alpha} \left[ \left(\hat l^z_{\alpha}\right)^2 -\frac 23 \right] Q_3   + \frac {B}2 \left( Q_2^2+Q_3^2 \right).
\label{HJT}
\end{eqnarray}
The generalized Kanamori parametrization\cite{Georges2013} was used to take into account correlation effects:
\begin{eqnarray}
\label{U}
 \hat H_{U} = (U-3J_H) \frac{\hat N (\hat N -1)}2 - 2J_H {\hat S}^2
 -\frac {J_H}2 {\hat L}^2 + \frac 52 \hat N. 
\end{eqnarray}
Here $J_H$ is the intra-atomic Hund's exchange, $U$ is the Hubbard intra-orbital Coulomb interaction, $\hat N$ is the total number of electrons operator. Note that the Hund's exchange interaction is written in a rotationally invariant form. Since we are dealing with an isolated ion case with fixed electron number,  one can safely put $U=0$.

Some limits such as the case of extremely strong SOC can be considered analytically, but for the general case of arbitrary electron filling and arbitrary ratio between different parameters -- most importantly $\lambda$, $g^2/B$ and $J_H$ -- one has to resort to numerical calculations, which were performed using exact diagonalization of~\eqref{H}.

%%%%%%%%%%%%%%%%%%%%%%%%%%%%%
\section{Results\label{sec:results}}
\subsection{Configuration: $\textbf d^1$\label{sec:results:d1}}  
\begin{figure}[t]
   \centering
  \includegraphics[width=0.51\textwidth]{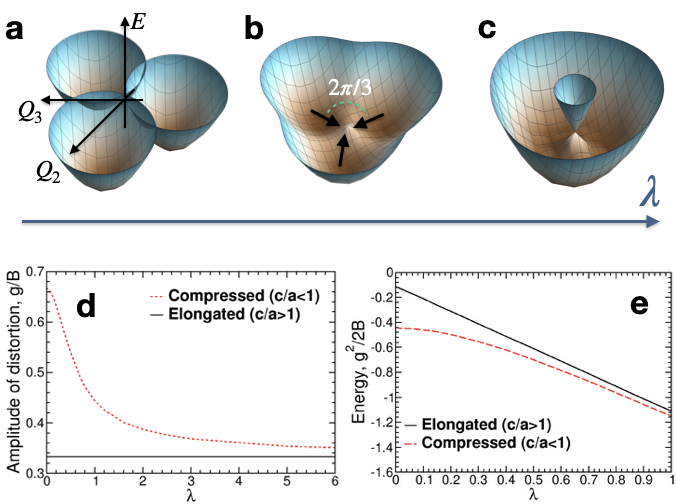}
  \caption{\label{Fig:energy-d1} Results for the $t_{2g}^1$ configuration ($g=B=1$). (a)--(c):~Evolution of the energy surface dependence on $Q_2$ and $Q_3$ modes as a function of the spin--orbit coupling constant~$\lambda$. (d):~Amplitude of distortion, $u$, as a function of $\lambda$ at the $(0, \pm u)$ points in the $Q_2Q_3$ plane. (e):~Energies of compressed and elongated octahedra as function of~$\lambda$.}
\end{figure}

We start with the simplest situation of a single $d$ electron on triply-degenerate $t_{2g}$ levels. Here one does not have to care about the Hund's coupling, and the treatment becomes rather straightforward. For $\lambda=0$ the maximal gain due to the JT effect is achieved if the electron occupies the low-lying $xy$ orbital, as shown in Fig.~\ref{Fig:CFS}a. This corresponds to a compression of ligand octahedra surrounding the transition metal, $c/a<1$. The total energy surface dependence on distortions can be readily obtained by diagonalizing \eqref{H} and is presented in Fig.~\ref{Fig:energy-d1}(a)--(c). For $\lambda=0$ it has three equivalent minima with energies $E_{JT}^0 = - \frac 49 \frac {g^2}{2B}$, corresponding to compressions along $x$, $y$, or $z$ directions\cite{Bersuker2006}.  The amplitude of the JT distortion in each of these minima is $u_{0} = \sqrt{ (Q_2^2+Q_3^2)} = \frac 23 \frac {g}{B}$. 

\begin{figure}[t]
   \centering
  \includegraphics
[width=0.45\textwidth]{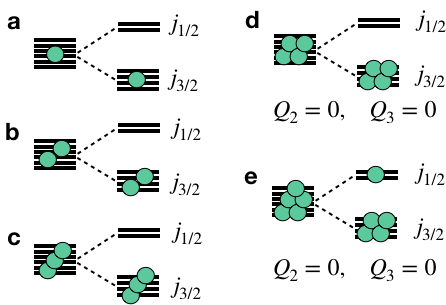}
  \caption{\label{Fig:SOC}  Level splitting diagrams due to spin--orbit coupling for different number of electrons, the case of the $jj$ coupling scheme (very large $\lambda$). }
\end{figure}

The electronic wavefunction corresponding to the paraboloid centered at $(0,-u_{\lambda=0})$ at the $Q_2Q_3$ plane has the $| xy \rangle = | l^z_0\rangle$ character. The increase of SOC leads to gradual admixture to this state (for which the SOC energy is zero) of the states $|l^z_{\pm 1} \rangle$ and to  a gradual suppression of the JT distortion, which, however, remains finite even for $\lambda \to \infty$. For very large $\lambda$ the amplitude of the JT distortion is reduced by a factor of~2, $u_{ \infty} =\frac 13 \frac g{B}$. More importantly, for very large $\lambda$ the distortions of both signs -- tetragonal compression $c/a<1$  and tetragonal elongation $c/a>1$ -- turn out to be equivalent. This becomes clear if we recall that for large $\lambda$ the electron would occupy one of the states of the $j_{3/2}$ quartet, Fig.~\ref{Fig:SOC}a. These states are
\begin{eqnarray}
\label{3/2-states}
|j_{3/2}, j^z_{3/2}\rangle &=& |l^z_1, \uparrow \rangle = - \frac 1{\sqrt 2} ( {\rm i} |xz, \uparrow \rangle + |yz, \uparrow \rangle),\nonumber \\
|j_{3/2}, j^z_{-3/2}\rangle &=& | l^z_{-1}, \downarrow \rangle 
=  -\frac 1{\sqrt 2} ({\rm i} |xz, \downarrow \rangle - |yz, \downarrow \rangle  ),
\nonumber \\
|j_{3/2}, j^z_{1/2}\rangle &=& \sqrt{\frac 2 3}|l^z_0, \uparrow \rangle + \frac 1 {\sqrt{3}}|l^z_{1}, \downarrow \rangle  \nonumber \\ 
&=&  \sqrt{\frac 2 3}|xy, \uparrow\rangle - \frac 1 {\sqrt{6}}|{\rm i}xz + yz, \downarrow \rangle, \nonumber \\
|j_{3/2}, j^z_{-1/2}\rangle &=& \sqrt{\frac 23}|l^z_0, \downarrow \rangle + \frac 1 {\sqrt{3}}|l^z_{-1}, \uparrow \rangle  \nonumber \\ 
\quad \quad &=&  \sqrt{\frac 2 3}|xy, \downarrow \rangle - \frac 1 {\sqrt{6}}| {\rm i}xz - yz , \uparrow \rangle. \label{eq:j32}
\end{eqnarray}

From \eqref{3/2-states} one sees that the state of Kramers doublet $|j_{3/2}, j^z_{\pm 3/2}\rangle$  is made from $l_z = \pm 1$ orbitals, i.e.\ only from the $xz$ and $yz$ orbitals. Using \eqref{HJT} one can readily find that the amplitude of the JT distortion for this doublet is $\frac 13 \frac gB$ and $E_{JT}^{\pm 3/2} = - \frac 19 \frac{g^2}{2B} = \frac 14 E_{JT}^0$. The states $|j_{3/2}, j^z_{\pm 1/2} \rangle$ consist predominantly of the $|l^z_0\rangle$ state, i.e.\ they contain a large fraction of the $xy$ orbital, for which the JT distortions (of opposite sign) are larger and the energy is lower. But there is also contribution coming from the $l^z_{\pm 1}$ orbitals in this doublet, see \eqref{eq:j32}, which causes the opposite distortion, and in effect the total distortion for the doublets   $|j_{3/2}, j^z_{\pm 1/2}\rangle$ turns out to be the same by magnitude but of opposite sign as compared to $|j_{3/2}, j^z_{\pm 3/2}\rangle$.  I.e.\ for the $d^1$ configuration strong SOC suppresses the JT distortion (by a factor of 2) and reduces the JT energy gain, $E_{JT}^{\lambda \to \infty} = E^0_{JT}/4$, and makes both tetragonal compression and tetragonal elongation equivalent. For intermediate values of $\lambda$ one expects that ``by continuity'' the state with $c/a<1$ would still lie lower than for $c/a>1$, but with increasing $\lambda$ they gradually approach each other. 

Our exact diagonalization calculation confirms this analysis: with increase of SOC three paraboloids corresponding to compressed octahedra join together, the edges formed by their crossings disappear, so that the energies of compressed and elongated octahedra turn out to be the same
at $\lambda \to \infty$, Fig.~\ref{Fig:energy-d1}(c). The distortion amplitudes (position of parabaloids) of compressed octahedra shown by the solid line in Fig.~\ref{Fig:energy-d1}d reduce smoothly down to $u_{\infty}$.

Two extra points should be mentioned here. The first one is that in our treatment, leading to Fig.~\ref{Fig:energy-d1}, we took into account the lowest order JT coupling and considered the lattice in the harmonic approximation, see~\eqref{HJT}. It has been shown that the inclusion of lattice anharmonicity and of higher order JT coupling leads to extra stabilization of the elongated octahedra, $c/a>1$ \cite{Kanamori1960,KhomskiiBrink2000}.  In effect the energy curve for $c/a>1$ in Fig.~\ref{Fig:energy-d1}(e), the black solid line there, would shift somewhat down and may cross the $c/a<1$ curve. I.e.\ depending on the parameters of the system such as SOC $\lambda$ and lattice anharmonicity, one may expect for $d^1$ with strong SOC both the JT contraction and JT elongation. This is nearly unthinkable for the JT effect in the case of $t_{2g}^1$ configuration without SOC, but apparently this is exactly what is observed experimentally for systems $A_2$TaCl$_6$, with $A=\rm K$, Rb, Cs~\cite{Ishikawa2019} .

The JT distortion of the ``usual'' sign, $c/a<1$, was observed for larger ions Rb, Cs, but the opposite distortion was found for the smaller $A$-ion K\null.  First, in these results we see that indeed the JT distortions are preserved even for the $5d$ ion Ta$^{4+}$ ($d^1$) with strong SOC\null.  Second, anharmonicity is stronger for the smaller ion K (which corresponds to a smaller tolerance factor\cite{khomskii2014transition}) and this can lead to the opposite JT distortions in K$_2$TaCl$_6$.

The second extra conclusion is potentially more important. As mentioned above, for doubly-degenerate $e_g$ orbitals ($e \otimes E$ problem) there exist important quantum effects, connected to the ``Mexican hat'' form of the energy surface, Fig. ~\ref{Fig:energy-d1}c, with (in the linear approximation) degenerate states (the trough in Fig.~\ref{Fig:energy-d1}c) for all mixing angles $\theta$ in $|\theta \rangle = \cos \theta\,|Q_3 \rangle + \sin\theta\,|Q_2 \rangle$  and especially with its conical intersection at the origin\cite{Bersuker1989}.

As discussed above, the energy surface of the $t \otimes E$ problem without SOC has the form shown in Fig.~\ref{Fig:energy-d1}(a); there is no strong quantum effects in this case. However, for $\lambda \to \infty$ we would deal with the $j_{3/2}$ quartet, Fig.~\ref{Fig:SOC}(a), which is actually {\it two} Kramers doublets, analogous to the doubly-degenerate $e_g$ electrons; thus it has, besides the Kramers degeneracy, an extra double degeneracy, naturally leading to the JT effect. The ``Mexican hat'' form of the energy surface for $e_g$ electrons, with the concomitant quantum effects, is well-known, see e.g.~\cite{Bersuker2006,khomskii2014transition}. The question arises whether we would recover in our case for strong SOC all the quantum effects typical for $e_g$ systems. And indeed this is what really happens. As is seen from Fig.~\ref{Fig:energy-d1}(a)--(c), with increasing $\lambda$ the energy surface gradually transforms from three paraboloids to the one of the Mexican hat. Thus, we see that for the $d^1$ configuration the increase of SOC, preserving the (weakened) JT effect, generates a novel situation with strong quantum effects.

However, actually the situation here, in many respects resembling that of $e_g$ electrons, is even richer and more interesting. Whereas in the usual JT problem  (without SOC which is quenched for $e_g$ electrons) orbitals and spins are completely decoupled and the ``Mexican hat'' describes purely orbital degrees of freedom, in our case, because of strong entanglement of spins and orbitals, when we move along the trough in the Mexican hat of Fig.~\ref{Fig:energy-d1}(c), the spin state and the total magnetic moment (its $z$-projection) also change.  This can lead to novel effects deserving special consideration. It would be very interesting to probe the corresponding effects experimentally.

The experimental situation for $4d$ and $5d$ systems is not quite clear. For systems  $M_2$TaCl$_6$, where $M = \rm K$, Rb, Cs (which can be viewed as double perovskites with ordered vacancies, e.g.\ K$_2$[Vacancy]TaCl$_6$), one indeed observes cubic--tetragonal structural transitions\cite{Ishikawa2019}, apparently driven by the JT effect. In some other cases the average bulk structure remains cubic, but local probes such as NMR or NQR detect a definite symmetry reduction of the same nature\cite{Lu2017}. The problem is also that it is often not easy to check the nature of distortions experimentally, since in addition to the JT distortions discussed in the present paper, in real materials there may be other types of distortions, such as rotations and tiltings of transition metal octahedra due to chemical pressure, characterized by the tolerance factor\cite{Goodenough}. As a result most materials having these structures turn out to be not cubic or tetragonal, but in most cases orthorhombic. In this sense it is better to work with systems with larger nonmagnetic ions (such as Rb, Cs in $A_2M$Cl$_6$), and the use of the local probes such as the one used in Ref.~\cite{Lu2017} to study these effects might be more promising.

\subsection{ Configuration: $\textbf d^2$} 
For the $d^2$ configuration and strong SOC one should in principle expect the JT effect still to be present -- as seen from Fig.~\ref{Fig:SOC}(b), for strong SOC we would have two electrons at fourfold degenerate $j=\frac32$ levels. In this case, however, we have to take into account the Hund's rule interaction of two electrons. Then with increasing SOC for $\lambda < J_H$ one would first have the situation with the LS coupling (we first form the total spin $S=1$, which couples with the orbital moment $L=1$ to form the total moment $J=2,1,0$), with the quintet $J=2$ lying lower. This state is still JT active. Such a state, with $S=1$, optimises the Hund's exchange, but is still not optimal for SOC.

For $\lambda > J_H$ we should use the $jj$ coupling scheme, now with two electrons on the $j_{3/2}$ quartet (Fig.~\ref{Fig:SOC}b) -- again, it seems, the typical JT situation. If these two electrons would be e.g.\ in $|j_{3/2}, j^z_{\pm 3/2} \rangle$  states, both with ``elongated’’ \eqref{lz1} orbitals, this would result in a JT distortion (local elongation, $c/a>1$). Similarly if both these electrons should occupy the states of the $|j_{3/2}, j^z_{\pm 1/2}\rangle$ doublet, these would also be JT active, with, as argued above, the JT distortion equal in magnitude but opposite in sign (compression, $c/a<1$). But, as is clear from the form of these wavefunctions~\eqref{eq:j32}, both these states are against the Hund's coupling: e.g.\ in the state $|j_{3/2}, j^z_{3/2}; j_{3/2}, j^z_{-3/2}\rangle$ the two electrons have opposite spin projections. But typically $J_H \gg E_{JT}$, thus one has, even for $\lambda>J_H$, to satisfy as much as possible the Hund's coupling, and such states, described above, turn out to be rather unfavourable. In the $jj$ scheme one can do better by putting one electron e.g.\ at the state $|j_{3/2}, j^z_{3/2}\rangle$, and another one at $|j_{3/2}, j^z_{1/2}\rangle$ (or both at the states $|j_{3/2}, j^z_{-3/2}\rangle$ and $|j_{3/2}, j^z_{-1/2}\rangle$). In this case we would gain maximum Hund's energy possible in the $jj$ scheme. However, these electron occupation would lead to the opposite JT distortions for two electrons, so that in the limit of large $\lambda$ they would cancel one another and the total JT distortion would asymptotically disappear. 
\begin{figure}[t]
   \centering
  \includegraphics[width=0.49\textwidth]{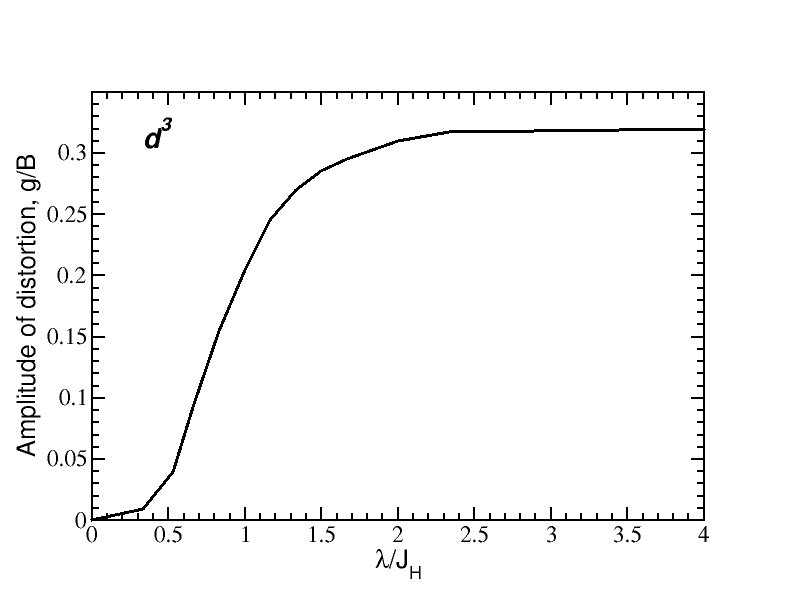}
  \caption{\label{Fig:energy-d3} The amplitude of the Jahn-Teller distortions for $t_{2g}^3$ configuration as a function of the ratio of the spin--orbit coupling strength~$\lambda$ and Hund's exchange~$J_H$, the results of the exact diagonalization calculations for $g=B=1$ and $J_H=0.5$.}
\end{figure}

Our numerical calculations shown in Fig.~S1 of Supplementary material\footnote{S.~Streltsov, D.~Khomskii, Supplementary materials}, indeed confirm this qualitative conclusion. Again, as in the $d^1$ case, for intermediate $\lambda$ the JT distortion would be still present, but (strongly) reduced; and again, ``by continuity’’,  the sign of it would be the same as for $\lambda =0$, i.e.\ local elongation $c/a>1$.  (In this case, in contrast to the $d^1$ situation, the anharmonicity would only support this distortion). But in contrast to the $d^1$ situation, in the case of $d^2$ the JT distortions will be {\it completely} suppressed at sufficiently strong SOC.

As mentioned above, the experimental situation is often complicated by the presence of other types of distortions (beside the JT ones). One of the materials with the transition metal having a $d^2$ configuration and which stays tetragonal down to very low temperatures is Sr$_2$MgOsO$_6$\cite{Yuan2015,Sarapulova2015}. The OsO$_6$ octahedra are elongated in accordance with our analysis. In order to study the effect of SOC we performed GGA${}+{}$U calculations for this materials and found that taking SOC into account suppresses the amplitude of the distortions by a factor of $\sim2$--3 (depending on the choice of interaction parameters and the type of the Hubbard correction)\footnote{E. Komleva and S. Streltsov, to be published}. These calculations support our results that SOC reduces the JT distortions in the case of $d^2$ configuration, but for $\lambda$ not too large they still remain finite.

\subsection{ Configuration: $\textbf d^3$} 
The situation with $d^3$ ions is especially interesting. In the limit of weak SOC we have a nondegenerate state $t_{2g}^3$ ($S=3/2$) without any orbital degeneracy and consequently without JT instability, Fig.~\ref{Fig:CFS}(c); but also with a completely quenched orbital moment, i.e.\ in this state we do not gain any SOC energy.  This is the classical situation with such ions as e.g.\ Cr$^{3+}$, Ru$^{5+}$ or Re$^{4+}$.

In the opposite limit, for very strong SOC, we would have three electrons, or one hole on the $j_{3/2}$ quartet (Fig.~\ref{Fig:SOC}(c)), so that the situation seems to be similar to the $d^1$ case in the large $\lambda$  limit.  Correspondingly, one might expect that in this limit the $d^3$ configuration would develop JT distortions. Our numerical results presented in Fig.~\ref{Fig:energy-d3} show that this is indeed the case: in the limit of $\lambda \to \infty$ the amplitude of the JT distortions is finite, $u \to \frac 13 \frac gB$. This is a rather unexpected and potentially very  important conclusion.
\begin{figure}[t]
   \centering
  \includegraphics
[width=0.45\textwidth]{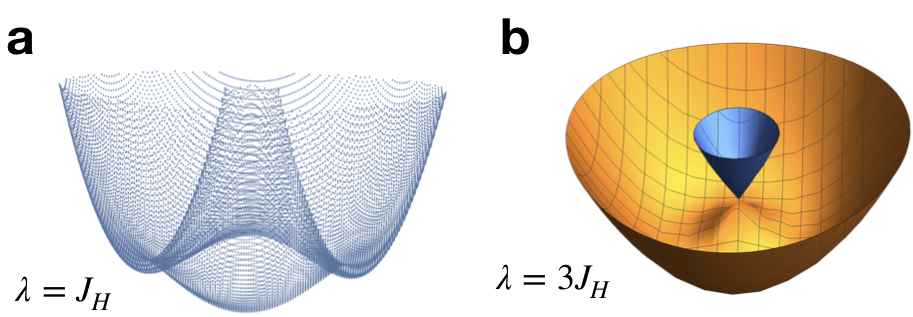}
  \caption{\label{Fig:d3-APES}  Configuration $d^3$. Energy surfaces for different ratios of $\lambda$ and $J_H$,  $g=B=1$. One may note three JT minima for $\lambda=J_H$ (corresponding to compressed octahedra), but already for $\lambda=3J_H$ they become nearly indistinguishable and the energy surface resembles the Mexican hat.}
\end{figure}

Detailed analysis shows that for small and intermediate $\lambda$ the compressed geometry is stabilized. Indeed, in order to gain maximum Hund's energy, it is better to put two electrons on the spin-orbitals with $j^z ={\pm\frac12}$ and the third electron on the one with $j^z = {+\frac32}$ (or $j^z = {-\frac32}$). But this state would immediately lead to a particular lattice distortion -- a compression of $ML_6$ octahedra (along $z$, $x$ or $y$-axes). However, with increasing $\lambda$ such an ``effective anharmonicity'' induced by the Hund's coupling becomes less and less important (compare Fig.~\ref{Fig:d3-APES} (a) and~(b)) and for $\lambda/J_H = 3$ the corrugation of the energy surface becomes practically indistinguishable and we restore the situation with the conical point and the Mexican hat, see Fig.~\ref{Fig:d3-APES}(b).  I.e.\ in this limit the system with one hole in the $j_{3/2}$ quartet behaves in the same way as the system with one electron, cf.\ Fig.~\ref{Fig:energy-d1}, although {\it a priori} this is not evident because of the possible role of the Hund's coupling in $d^3$ case vs.\ $d^1$.

A good system on which one could study the situation with $d^3$ configuration is for example K$_2$ReCl$_6$. Structural, thermodynamic and magnetic measurements demonstrate that there exist in this system not one but three structural transitions\cite{Busey1961,Busey1962,Lynn1978a}. Their origin is not yet clear. Some of them could be mainly connected with the rotation and tilting of ReCl$_6$ octahedra as explained above. But we suppose that the eventual JT character of Re$^{4+}$ ($d^3$) ions, activated by strong SOC, could also be involved in at least one of these transitions. It would be very interesting to check our conclusions on similar materials with Rb, Cs instead of K for which there are less chances to have rotation and tilting of ReCl$_6$ octahedra.

\subsection{ Configurations: $\textbf d^4$ and $\textbf d^5$}
The situation for $d^4$, and also for $d^5$ configurations is in a some sense simpler. Without SOC there should exist in both of them the JT distortions of opposite signs, see Fig.~\ref{Fig:CFS}d,e. On the other hand, for strong SOC $d^4$ ions would become $J=0$ singlets, both in the LS ($\lambda < J_H$) and $jj$ ($\lambda>J_H$) coupling  schemes, see Fig.~\ref{Fig:SOC}d, i.e. there would be of course no JT distortion in this state.
\begin{figure}[t!]
   \centering
  \includegraphics[width=0.49\textwidth]{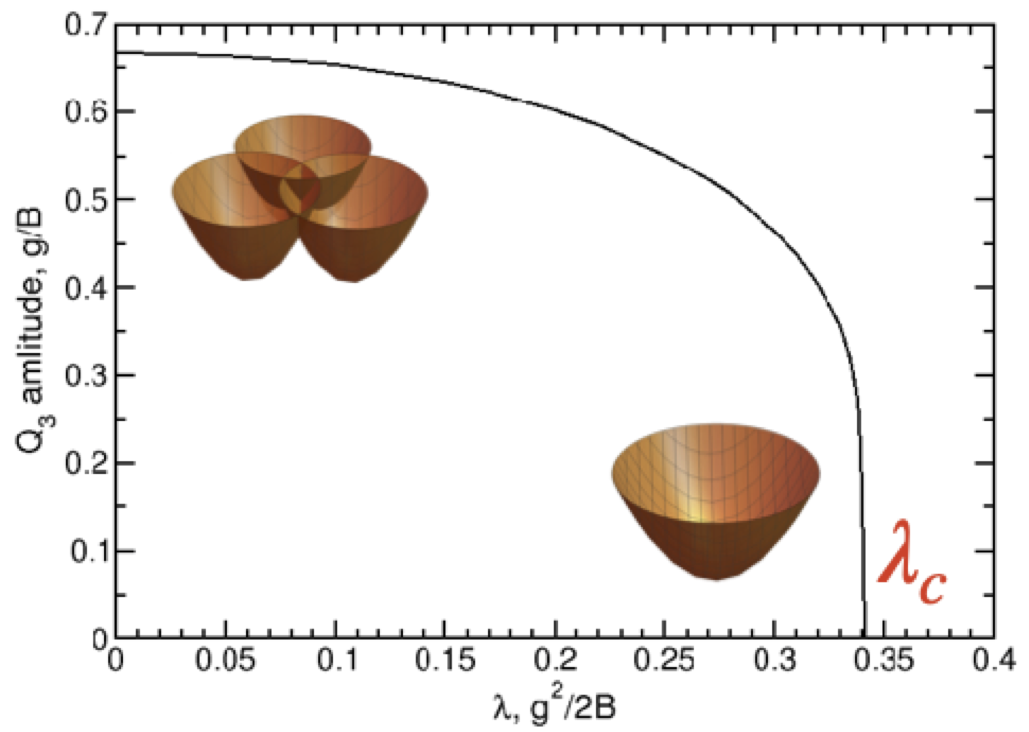}
  \caption{\label{Fig:energy-d5} The amplitude of the Jahn-Teller distortions for $t_{2g}^5$ configuration as a function of spin--orbit coupling strength~$\lambda$. Results of the exact diagonalization calculations for $g=B=1$. Insets demonstrate how the total energy surface as a function of the $Q_2$ and $Q_3$ distortions evolves with~$\lambda$.}
\end{figure}

The same is true also for the $d^5$ case (to which the popular Ir$^{4+}$ belongs): for strong SOC it is a Kramers doublet $J=\frac12$, of course without JT instability, see Fig.~\ref{Fig:SOC}(e). The question is only how the JT effect disappears with increasing~$\lambda$.

Our calculations, the results of which are presented in Fig.~\ref{Fig:energy-d5} and Fig.~S2 of Supplementary materials, show that in both cases the JT effect disappears above some critical value~$\lambda_c$ in an almost abrupt way. Thus, in contrast to the $d^2$ case, for which the JT distortions gradually subside with increasing $\lambda$ and disappear at large~$\lambda$ only asymptotically, Fig.~S1, for both $d^4$ and $d^5$ configurations the JT effect is strictly absent for~$\lambda>\lambda_c$.

%%%%%%%%%%%%%%%%%%%%%%%%%%%
\section{Intersite Jahn-Teller coupling and spin--orbit interaction}

When one deals not with isolated JT centres (as was done in Sec.~\ref{sec:results}) but with solids, various additional factors may affect the result of interplay between JT effect and spin--orbit coupling. First of all, there are effects related to the formation of electronic bands\cite{Kaplan1995}; second, the JT centres may interact via the field of phonons\cite{Gehring1975}, and last but not least there is an exchange mechanism, which couples transition metal ions whose $d$ bands are not filled completely\cite{KK-UFN}. The result also strongly depends on the geometry. Thus, this is a rather general and complicated problem to take into account all these factors, and its treatment lies far beyond the scope of the present paper. But we would like to demonstrate on a particular example that the effects found in previous sections can survive even in concentrated systems.
\begin{figure}[t!]
   \centering
  \includegraphics[width=0.49\textwidth]{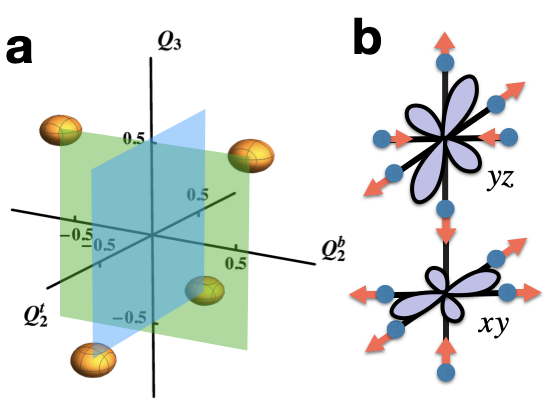}
  \caption{\label{Fig:CJT1} (a)~Four solutions corresponding to the total energy minimum of a pair of Jahn-Teller centres in a common corner geometry ($t_{2g}^1$ configuration, $\lambda=0$). $Q_3$, $Q_2^u$, and $Q_2^t$ are given in units of~$g/B$. (b)~Distortions (and orbitals) corresponding to one of these minima.}
\end{figure}

We consider two JT ions having, for simplicity, just a single electron in the $t_{2g}$ shell. Let these ions have octahedral surroundings and share a common corner lying along the $z$-direction, see Fig.~\ref{Fig:CJT1}(b). Because of the presence of a common middle ligand in such geometry the $Q_3$ mode for the bottom ion, $Q_3^b$, turns out to be coupled with the mode of the same symmetry, $Q_3^t$, of the upper (top) ion. Thus, when we fix the volume, i.e.\ the distance between the top and bottom JT ions, the elongation of the bottom octahedra will automatically result in the compression of the upper one (and vice versa). Therefore we would have $Q_3$ distortions of the two octahedra which are the same in magnitude but opposite in sign, and \ in the expressions for $\hat H_{JT}$ this distortion must enter with opposite signs for upper and lower ions (the second term in Eq.~\eqref{HJT}).  Thus in this geometry one should take $Q_3^t = -Q_3^b =Q_3$, while we do not apply any restrictions on $Q_2^b$ and  $Q_2^u$ and they can be different. (Note, however, that in a crystal having a lattice of such centres all distortions must of course be ``consistent'' with each other).

The resulting Hamiltonian for such a cluster with two electrons and two JT centres interacting only via elastic coupling can be again solved numerically by the exact diagonalization technique (in these calculations we neglect Hund's coupling, but include large~$U \gg \{ g,B, \lambda \}$ to avoid the situation with double occupancy). While each non-interacting JT centre has three minima in the adiabatic potential energy surface without spin--orbit coupling (Fig.~\ref{Fig:energy-d1}), two coupled octahedra have four minima. These minima are at $(0, \pm \frac 1{\sqrt 3}, \frac 12)$ and $(\pm \frac 1{\sqrt 3}, 0, -\frac 12)$ points in the $Q_2^b, Q_2^t, Q_3$ space, see Fig.~\ref{Fig:CJT1}(a). These points correspond to the situation when, e.g., at the lower site the octahedron is compressed in the $z$-direction and the electron occupies the $xy$ orbital, while at the upper site we have the octahedron compressed along the $x$-direction with the electron on the $yz$ orbital (as shown in Fig.~\ref{Fig:CJT1}(b)), or compressed along the $y$-direction with the electron on the $xz$ orbital. Or, vice versa, the $xy$ orbital is occupied at the upper site, and at the lower one we have the electron sitting on the $yz$ or $xz$ orbitals; this gives in effect four minima. Such local distortions allow us to gain as much local JT energy as possible at each site, which for one electron prefer compressed octahedra (in the $z$, $x$ or $y$-directions), see Fig.~\ref{Fig:CFS}(a), and at the same time to minimize the total strain in the crystal, cf.~\cite{Khomskii2001,Khomskii2003}. Still, the coupling between sites in this regime leads to a deviation from the absolute energy minima of isolated JT centres:  the energy of two coupled JT centers $E^{\rm coupled}_{JT}=-\frac 56 \frac {g^2}{2B}$ is larger than the energy of two isolated such centers $E^{\rm 2-is.}_{JT}=-\frac 89 \frac {g^2}{2B}$. Thus, the coupling shifts us from the minima of paraboloids of Fig.~\ref{Fig:energy-d1}(a), i.e.\ the interaction between sites partially competes with the local JT effect.

As has been discussed in Sec.~\ref{sec:results:d1}, spin--orbit coupling suppresses any difference between compressed and elongated octahedra, so that one expects that the competition between the on-site and intersite effects mentioned above should disappear. Indeed, in the limit of $\lambda \to \infty$ the Mexican hat geometry of the energy surface appears, but now we have degenerate solutions both in the $Q_3Q_2^b$ and $Q_3 Q_2^t$ planes (in Supplemental materials, Fig. S3, we show the resulting energy surface in $Q_3Q_2^b$ for $\lambda=5$ taking $Q_2^t$ as a constant).

\begin{figure}[t!]
   \centering
  \includegraphics[width=0.49\textwidth]{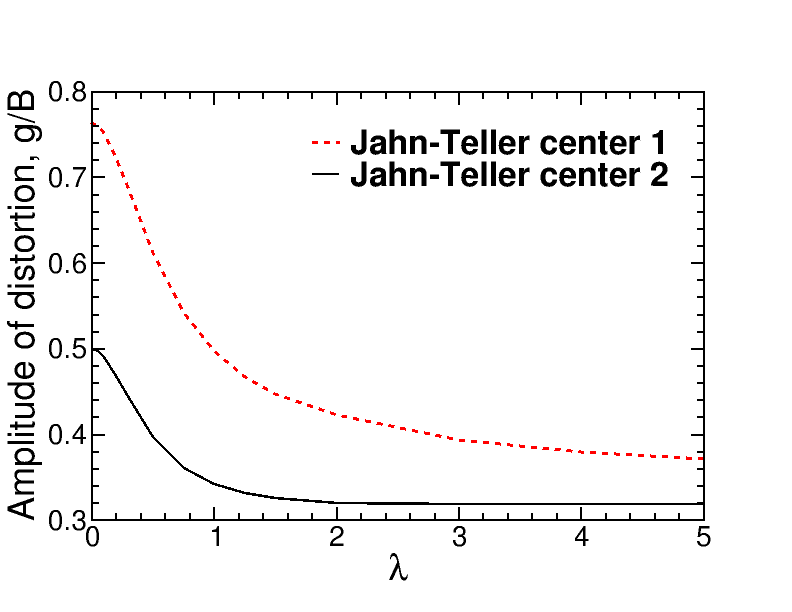}
  \caption{\label{Fig:CJT2} The amplitude of the distortion for a pair of Jahn-Teller centres as a function of $\lambda$ ($g=B=1$).}
\end{figure}

The dependences of the amplitude of JT distortion for both JT centres (top and bottom) is presented in Fig.~\ref{Fig:CJT2}. One may see that spin--orbit coupling still partially suppresses the Jahn-Teller effect even if we take into account elastic coupling between octahedra (though one might notice that this effect turns out to be somewhat weaker; compare Fig.~\ref{Fig:energy-d1}(d) and Fig.~\ref{Fig:CJT2}).  Note that one sees here the effect mentioned above: for strong spin--orbit coupling the local and intersite JT distortions do not compete anymore, the magnitude of JT distortions at each site approaches the same value as for isolated sites, $\sim \frac 13 \frac gB$, cf.\ Fig.~\ref{Fig:energy-d1}(d) and Fig.~\ref{Fig:CJT2}, and the corresponding minimum energy of coupled sites is close to twice the energy of a single isolated site in this limit, i.e.\ $E^{\lambda \to \infty}_2  \approx - \lambda - \frac 2 9 \frac{g^2}{2B}$.

%%%%%%%%%%%%%%%%%%%%%%%%
\section{Conclusions} 

We start this section by pointing out that the approach used in the paper is rather different from a ``classical'' way of treating the Jahn-Teller effect in the presence of the spin--orbit coupling. Typically one starts with one-electron description neglecting strong electron correlations, but treating both the vibronic coupling and the spin-orbit coupling perturbatively~\cite{Bersuker2006}. The present approach is very different, it is based on numerically exact solution of the {\it many-electron} problem taking into account the spin--orbit coupling with the chosen form of the vibronic coupling, see \eqref{H}-\eqref{U}. This allows us to treat within the same calculation scheme correlation effects, spin--orbit and vibronic coupling. 

In the present paper we demonstrate that there exists a very nontrivial interplay between spin--orbit coupling and the Jahn-Teller effect already at the single-site level. In most cases ($d^1$, $d^2$, $d^4$, $d^5$ configurations in the low-spin state typical for $4d$ and $5d$ ions) spin--orbit coupling counteracts and suppresses Jahn-Teller distortions -- gradually for $d^1$ and $d^2$ cases, but almost abruptly for $d^4$ and $d^5$. It still survives for the $d^1$ occupation, though it is reduced by a factor of~2, whereas for the $d^2$ case it completely disappears for large SOC\null. However there exists also the opposite effect: for the $d^3$ configuration, which does not have orbital degeneracy for $\lambda=0$, the spin--orbit coupling {\it activates} the Jahn-Teller effect and these Jahn-Teller distortions may become substantial for strong spin--orbit coupling. This could make the behaviour of $4d$ and $5d$ systems with $d^3$ occupation much richer than those based on $3d$ elements.
 
Also, whereas the Jahn-Teller effect ``survives’’  for the $d^1$ configuration at strong spin--orbit coupling, different types of  Jahn-Teller distortions, elongation and compression of metal--ligand octahedra, become almost degenerate, so that the relatively weak extra factors, such as lattice anharmonicity, can finally determine the detailed type of the Jahn-Teller distortions. Moreover, in this paper we considered the JT coupling of degenerate $t_{2g}$ electrons only with the $E_g$ distortions (tetragonal and orthorhombic). However, the $t_{2g}$ levels can also be split by trigonal distortions, triply-degenerate $T_{2g}$-vibrations (the $t \otimes T$ problem).  The vibronic coupling constants for $T_{2g}$ modes can be even larger than for $E_g$~\cite{Iwahara2018}. This $t \otimes T$ problem in the presence of the spin--orbit coupling for different electron number is much more complicated and less transparent (e.g.\ one cannot easily draw the energy surface in four-dimensional space). Still, in the Jahn-Teller physics there are some methods developed for the treatment of the $t \otimes T$ problem. Our preliminary results demonstrate that the mutual influence of the Jahn-Teller effect and spin--orbit coupling is qualitatively similar for the coupling to trigonal distortions to that for tetragonal and orthorhombic ones considered in this paper. We are planning to address this situation in more details in future. And of course generally one needs to solve the full $t \otimes (E+T)$ problem. 

In considering the coupled Jahn-Teller and spin--orbit interactions we paid main attention to the modification of the Jahn-Teller distortions by the spin--orbit coupling. But of course there exist also the opposite, inverse effect: the suppression of the spin--orbit coupling by Jahn-Teller distortions and by vibronic effects. This is already clear from the Fig.~\ref{Fig:CFS}: unquenched orbital moments existing e.g. for one or two electrons per site without Jahn-Teller distortions (for triply-degenerate $t_{2g}$ levels), leading to finite spin--orbit coupling,  are quenched due to Jahn-Teller distortions, Fig.~\ref{Fig:CFS}(a,b). I.e. in this limit Jahn-Teller distortions strongly suppress the spin--orbit coupling. This effect persists also for larger values of the spin--orbit coupling. It can be thought of as a result of the suppression of nondiagonal matrix elements by vibronic effects, the so-called Ham reduction factor~\cite{Ham1965}.  The reduction of effective spin-orbit splitting due to the conventional or pseudo-Jahn-Teller effect was indeed seen in some experiments and calculations, e.g. very recently in defects in diamond~\cite{Ciccarino2020}.

Besides the very existence of the Jahn-Teller effect in the case of strong spin--orbit coupling, we have shown that significant quantum effects, absent for triple $t_{2g}$ degenerate states without spin--orbit coupling, would appear at finite spin--orbit coupling. In that sense this  situation becomes very similar to that of the doubly-degenerate $e_g$ case: the energy surface of the Mexican hat type is formed, with the real singularity -- conical intersection, which is known to lead to many important physical consequences in spectroscopy, in dynamics etc. Actually here the situation is even richer because of the spin--orbit entangled character of electronic states for strong spin--orbit coupling.

When going to concentrated systems, also intersite effects -- notably coupling via phonons and superexchange interaction -- start to play an important role. We have seen that in the case of a single $t_{2g}$ electron the coupling via elastic terms does not change the situation for strong spin--orbit coupling qualitatively: as was the case for isolated sites, the JT distortion survives, albeit weakened for strong SOC\null. Nevertheless some interesting changes occur in this case: whereas for weak SOC the local and intersite JT effects partially compete, strong SOC ``relieves'' this competition, and the JT effect can work at full force (of course weakened by SOC but not by intersite coupling). However, generally the intersite effects can significantly influence the behaviour of the system. Thus intersite effects can lead to a more complicated form of exchange interaction\cite{Chen2010a,Chen2011} or can give a magnetic state for $d^4$ systems with the nominally singlet $J=0$ ions (singlet magnetism), see \cite{Khaliullin2013,khomskii2014transition}. Consequently, the question of Jahn-Teller activity of such ions could again become relevant (due to the effective admixture to the $J=0$ state of the other, potentially  Jahn-Teller active configurations). But in any case, even for concentrated systems the effects considered in this paper should be included and should serve as a background, the fundament on which the building of bulk solids should be based.

\section{Acknowledgements}
We are grateful to G.~Jackeli, L.~Chibotaru and G.~Khaliullin for various stimulating discussions and to the UK consulate in Ekaterinburg for organizing the ``Scientific cafe'' at Institute of metal physics in 2019, where this work was initiated. This research was carried out within the state assignment of FASO of Russia via program ``Quantum'' (No.\ AAAA-A18-118020190095-4).  We also acknowledge the support by the Russian Foundation for Basic Researches (RFBR 20-32-70019) and the Russian Ministry of Science via contract 02.A03.21.0006 and the Deutche Forschungsgemeinschaft (DFG, German Reseach Foundation), project number 277146847-CRC 1238.

\bibliography{../library}

\end{document}